\documentstyle[prb,aps]{revtex}
\twocolumn
\begin{document}

\draft
\twocolumn[\hsize\textwidth\columnwidth\hsize\csname@twocolumnfalse\endcsname

\title{Evidence against strong correlation in 4$d$ transition metal oxides,
CaRuO$_3$ and SrRuO$_3$}

\author{Kalobaran Maiti\cite{kbm} and Ravi Shankar Singh}
\address{Department of Condensed Matter Physics
and Materials Science, Tata Institute of Fundamental Research, Homi Bhabha Road, Colaba,
Mumbai - 400 005, India}

\date{\today}
\maketitle

\begin{abstract}

We investigate the electronic structure of 4$d$ transition metal
oxides, CaRuO$_3$ and SrRuO$_3$. The analysis of the photoemission
spectra reveals significantly weak electron correlation strength
($U/W\sim$0.2) as expected in 4$d$ systems and resolves the long
standing issue that arose due to the prediction of large $U/W$
similar to 3$d$-systems. It is shown that the bulk spectra,
thermodynamic parameters and optical properties in these systems
can consistently be described using first principle approaches.
The observation of different surface and bulk electronic
structures in these weakly correlated 4$d$ systems is unusual.

\end{abstract}


]

Hubbard model consisting of inter-site hopping energy, $t$
($\propto W$=bandwidth) and on-site Coulomb repulsion energy, $U$
has been widely accepted as the simplest model to capture
correlation effects in the electronic structure. Thus, effective
electron correlation strength can be described by a single
parameter, $U/W$ ($U/W>1\Rightarrow$insulator,
$U/W<1\Rightarrow$metal). Correlation effects in 3$d$ transition
metal oxides (TMO) have extensively been studied using this model
during the past few decades following the discovery of many exotic
properties in these systems.\cite{RMP} Despite simplicity of the
model, the major difficulty arises from large $U$ leading to
non-perturbative nature of the problem, and the presence of
various competing effects such as different types of long-range
order, interplay between localization and lattice coherence etc.

4$d$ orbitals in 4$d$ TMOs are more extended than 3$d$ orbitals in
3$d$ TMOs. Thus, correlation effect is expected to be less
important in these systems and provide a suitable testing ground
for the applicability of various {\it ab initio} approaches. This
is reflected by only a small mass enhancement\cite{cao,allen}
($m^\star/m_b$ = 3.0 in SrRuO$_3$; $m^\star$ = effective mass,
$m_b$ = band mass) in the specific heat measurements. However,
photoemission\cite{fujimori,park} and optical\cite{ahn} data
reported so far are significantly different from their {\it ab
initio} results. All these studies predict a large $U/W$ similar
to that observed in 3$d$ TMO in contrast to the behavior expected
from the highly extended nature of 4$d$ systems. It is thus
believed that various approximations (slave boson, techniques in
the limit of infinite dimensions, exact calculations for finite
size system {\it etc.}) are necessary to simulate the material
properties of these systems. In addition, tetravalent ruthenates
have drawn significant attention recently due to the discovery of
interesting magnetism,
\cite{cao,allen,fujimori,park,ahn,mazin,cox-fukunaga-rrao}
unconventional superconductivity,\cite{sr2ruo4} metal-insulator
transitions,\cite{tl2ru2o7} {\it etc.}

In this paper, we present evidence for weak correlations among
4$d$ TMOs. We report the results of our investigation of the
electronic structure of 4$d$ TMOs, SrRuO$_3$ and CaRuO$_3$.
SrRuO$_3$ is a ferromagnetic metal. Isostructural and
isoelectronic, CaRuO$_3$ (with slightly different Ru-O-Ru bond
angle; 150$^o$ in CaRuO$_3$ and 165$^o$ in SrRuO$_3$) exhibits
antiferromagetic behavior.\cite{cao,rss} Photoemission spectra at
different photon energies reveal different electronic structures
for surface and bulk in {\em both} the cases. It is to note here
that early 3$d$ TMOs as well as rare-earth systems are known to
exhibit different surface and bulk electronic
structures,\cite{casrvo,lacavo,kbmprb,RE} which was attributed to
an enhancement of $U/W$ at the surface compared to that in the
bulk. It was necessary to employ the dynamical mean field
theoretical approach (DMFT) within the limit of infinite
dimensional Hubbard model to determine the bulk electronic
structure in these 3$d$ systems.\cite{liebsch-andersen}
Interestingly, the bulk spectra in this study could be described
remarkably well using {\em ab initio} approaches. Present results,
thus, resolve three fundamental issues in these systems. (a)
Electron correlation is significantly weak in 4$d$ TMOs ($U/W
\sim$ 0.2). (b) First principle approaches are sufficient to
simulate the bulk photoemission and optical responses, and to
reproduce $m^\star/m_b$ obtained from the specific heat
measurements. (c) The surface and bulk electronic structures can
be different even in weakly correlated systems.

High quality polycrystalline samples (large grain size achieved by
long sintering at the preparation temperature) were prepared by
solid state reaction method using ultra-high purity ingredients
and characterized by $x$-ray diffraction (XRD) patterns and
magnetic measurements as described
elsewhere.\cite{cao,fujimori,rss} Sharp XRD patterns reveal pure
GdFeO$_3$ structure with similar lattice constants as observed for
single crystalline samples.\cite{cao} Magnetic susceptibility
measurements exhibit a ferromagnetic transition in SrRuO$_3$ at
165~K and the signature of antiferromagnetic interactions in
CaRuO$_3$ at 180~K. The magnetic moment of 2.7~$\mu_B$ in
SrRuO$_3$ and 3~$\mu_B$ in CaRuO$_3$ in the paramagnetic phase is
close to their spin-only value of 2.83~$\mu_B$ for
$t_{2g\uparrow}^3 t_{2g\downarrow}^1$ configurations at Ru sites.
Photoemission measurements were performed on {\it in situ}
(4$\times$10$^{-11}$~torr) scraped samples\cite{note1,note2} using
SES2002 Scienta analyzer at room temperature (paramagnetic phase)
in order to avoid complications due to different long-range
orders. The experimental resolution was 7 meV, 0.8 eV and 0.9 eV
for measurements with monochromatic He~{\scriptsize II},
Mg~$K\alpha$ and Al~$K\alpha$ lines.

Valence band spectra obtained at He~{\scriptsize II}, Mg~$K\alpha$
and Al~$K\alpha$ excitation energies are shown in Fig. 1. The
signature of three discernible features is evident in the figure.
While the features appearing at binding energies $>$~2.5 eV are
large in He~{\scriptsize II} spectra, the relative intensity
between 2.5-0 eV binding energies is significantly enhanced in the
$x$-ray photoemission (XP) spectra. Considering strong dependence
of the relative transition matrix elements on excitation
energies,\cite{yeh} the 0-2.5~eV feature can be attributed to
essentially Ru 4$d$ electron excitations with the O~2$p$
contributions appearing beyond 2.5~eV.

In order to understand these results, we calculated the electron
density of states (DOS) of CaRuO$_3$ and SrRuO$_3$ using full
potential linearized augmented plane wave method (FLAPW) within
the local density approximations (LDA) for the experimentally
obtained crystal structure in the paramagnetic phase.\cite{wien}
The convergence was achieved using 512 $k$-points within the first
Brillouin zone. Total DOS (TDOS) and partial DOS (PDOS)
corresponding to Ru 4$d$ and O 2$p$ states are shown in Fig. 2. No
significant contribution from Ca and Sr states are observed in
this energy range. Three distinct groups of peaks are evident in
the figure in the occupied part. Non-bonding O 2$p$ states appear
between 2-4 eV and the bonding states with dominant O 2$p$
contributions appear at higher binding energies. This is also
manifested in Fig. 1 by an enhancement in intensity at 7~eV in the
XP spectra compared to the He~{\scriptsize II} spectra.
He~{\scriptsize II} spectra exhibit a peak below 6~eV. The
intensity at $E_F$ arises primarily due to Ru 4$d$ split $t_{2g}$
bands. The broad and empty $e_g$ bands appear beyond 1~eV above
$E_F$.

Small distortion of RuO$_6$ octahedra lifts the degeneracy of the
$t_{2g}$ band as shown in the insets of Fig. 2. While $d_{xy}$,
$d_{yz}$ and $d_{xz}$ bands in SrRuO$_3$ have similar width ($W$ =
2.4~eV), the total bandwidth is about 3.1~eV due to the splitting
of $t_{2g}$ band. In CaRuO$_3$, the total bandwidth is about
2.4~eV. Interestingly, the width of $d_{xy}$, $d_{xz}$ and
$d_{yz}$ bands varies between 2.3-2.4~eV which is similar to those
in SrRuO$_3$. {\em This reveals that the small change in Ru-O-Ru
bond angle between SrRuO$_3$ and CaRuO$_3$ has essentially no
influence on $W$ (the bandwidth of the individual d-bands) and
thus, $U/W$ in these systems.} The center of mass for $d_{xy}$
band appears very close to $E_F$ in {\it both} the cases, while
$d_{xz}$ and $d_{yz}$ bands appear at lower energies.

It is now clear that the feature close to $E_F$ has primarily
Ru~4$d$ character. High resolution in the He~{\scriptsize II}
spectra leads to a minimal overlap between O~2$p$ and Ru~$t_{2g}$
bands (see Fig. 1). Hence, it is possible to delineate Ru~$t_{2g}$
contributions quite reliably by subtracting O~2$p$ tail as shown
in the inset (I) of Fig. 1 for SrRuO$_3$. Ru $t_{2g}$ bands, thus
extracted for {\em both} CaRuO$_3$ and SrRuO$_3$ are shown in Fig.
3(a). Large intensity at $E_F$ represents the extended states as
also seen in the band structure calculations and is commonly known
as 'coherent feature'. Interestingly, the maximum intensity
appears around 1.2~eV as observed before,\cite{fujimori,park}
while the theoretical intensity peaks at about 0.5 eV with
negligible contributions beyond 1 eV (see Fig. 3(b)). Thus, this
feature is often attributed to the signature of the electronic
states essentially localized due to electron correlations and is
termed as 'incoherent feature'. Spectral modifications observed in
Fig. 3(a) indicate larger correlation effects in CaRuO$_3$
compared to SrRuO$_3$ as predicted from the structure. While large
intensity at $E_F$ indicates metallic character, the complete
dominance of this apparent incoherent feature was taken as
evidence of the presence of strong correlation effects in previous
studies.\cite{fujimori,park,ahn} We present our results below
establishing that this feature essentially arises due to the
contributions from the surface electronic structure.

Interestingly, the XP spectra in Fig. 1 exhibit highest intensity
at 0.5 eV for the Ru 4$d$ band even though the intensities at
higher energies contain contributions from resolution broadened O
2$p$ band-tails. In order to subtract these contributions from the
XP spectra, we broadened the peaks 1 and 2 in inset (I)
representing the O 2$p$ features in the He~{\scriptsize II}
spectra upto the resolution broadening of the XP measurements and
the relative intensities of the features 1 and 2 were obtained
{\em via} least-square-error-fit. The subtracted spectra shown in
Fig. 3(c) clearly possess different lineshape compared to the
solid lines representing resolution broadened He~{\scriptsize II}
spectra. The $x$-ray photoelectrons has significantly larger
escape depth compared to the ultra-violet photoelectrons, which
leads to a significantly larger bulk sensitivity in XPS compared
to that in UPS. Thus, it is obvious that the bulk and surface
electronic structures are significantly different in these
systems. {\em We do not observe any change in lineshape of the
$t_{2g}$ band in the XP spectra of CaRuO$_3$ and SrRuO$_3$ as
expected from the band structure results described above.}

Photoemission intensity can be expressed as $I(\epsilon) =
[1-e^{-d/\lambda}]f^s(\epsilon) + e^{-d/\lambda}f^b(\epsilon)$,
where $d$ is the thickness of the surface layer and $\lambda$ is
the escape depth of the photoelectrons. $f^s(\epsilon)$ and
$f^b(\epsilon)$ represent the surface and bulk spectral functions,
respectively. Since $d$ and $\lambda$ are expected to change in
the same way due to the difference between 3$d$ and 4$d$ sites,
experimentally estimated $d/\lambda$-values in a similar system
CaSrVO$_3$\cite{casrvo} are a good approximation to extract
$f^s(\epsilon)$ and $f^b(\epsilon)$ in these systems. We observe
that a change in $d/\lambda$ values by more than 10\% leads to
unphysical intensities providing confidence in this procedure. The
extracted $f^s(\epsilon)$ and $f^b(\epsilon)$ are shown in Fig. 4.
$f^s(\epsilon)$ in both the cases is dominated by the intensity
centered at 1~eV. $f^b(\epsilon)$ exhibit large coherent feature
with the peak at about 0.5 eV as observed in the {\it ab initio}
calculations. The feature around 2 eV indicates the presence of
some degree of correlation effects, which is estimated below.

The spectral function can be expressed as, $f^b(\epsilon)=-{1\over
\pi} Im \sum_k G_k(\epsilon)$, where $G_k(\epsilon)$ is the
retarded Green's function representing the many electron system
and is given by
$G_k(\epsilon)=1/(\epsilon-\Sigma_k{(\epsilon)}-\epsilon_k)$.
$\Sigma_k{(\epsilon)}$ is the self energy of the system. For small
$U$, $\Sigma_k{(\epsilon)}$ can be calculated using perturbation
approach upto the second order term\cite{treglia} within the local
approximations. We have used Ru $t_{2g}$ PDOS for these
calculations. The calculated spectral functions are shown by solid
lines in Fig. 4. The fit to the experimental spectra is
remarkable. Most interestingly, $U/W$ is found to be significantly
small (0.24 for CaRuO$_3$ and 0.21 for SrRuO$_3$; $U = 0.6 \pm
0.05$~eV in both the cases) in sharp contrast to all previous
predictions.\cite{fujimori,park,ahn} A small variation in $U/W$
leads to significantly large spectral weight transfer as shown in
Fig 4(a). We calculated the mass enhancement factor following the
relation, ${m^\star\over m_b} = 1
-
{{\partial\over{\partial\epsilon}}}\left(Re\Sigma(\epsilon)\right)|_{\epsilon
= E_F}$. Interestingly, $m^*/m_b$ for SrRuO$_3$ is almost the same
(= 2.9) as that obtained (= 3.0) from specific heat measurements.
While $U/W$ in CaRuO$_3$ is little larger than that in SrRuO$_3$,
$m^*/m_b$ in CaRuO$_3$ is found to be close to SrRuO$_3$ and
somewhat smaller than its experimental value.

It is important to realize here that the optical response
(particularly in transmission mode) corresponds essentially to the
bulk electronic structure.\cite{ahn} Thus, the description of weak
correlation effects in the bulk electronic structure should also
be reflected in the optical spectra. We, thus, calculated the
optical response by calculating the joint density of states using
Wien-program.\cite{wien} The calculated spectra (correlation
effects are not considered) shown in the inset of Fig. 4(b)
resemble the experimental spectra by Ahn {\em et al.} \cite{ahn}
remarkably well. {\em This study, thus, establishes for the first
time that these weakly correlated 4$d$ TMOs provide a model
systems for the applicability of the first principle approaches in
determining the thermodynamic, optical and spectroscopic
properties.}

Differences in the surface and bulk electronic structures in 3$d$
TMOs as well as in rare earths were attributed to the enhancement
of $U/W$ due to the band narrowing at the
surface.\cite{casrvo,liebsch-andersen} Thus, the observation of
different $f^s(\epsilon)$ and $f^b(\epsilon)$ in these weakly
correlated 4$d$ TMOs as well is unusual. Most surprisingly, the
surface peak appears at lower binding energies than the incoherent
peak in the bulk spectra. It is observed that small distortion of
RuO$_6$ octahedra in the orthorhombic structure lifts the
degeneracy of the bulk $t_{2g}$ band (see insets in Fig. 2).
Whether the surface layer consists of Sr/Ca-O or Ru-O layer, the
absence of periodicity along the surface will enhance this
distortion leading to a different crystal field symmetry
presumably close to $D_{4h}$ symmetry ($t_{2g}\Rightarrow e_g +
b_{2g}$); where, $d_{xz}$ and $d_{yz}$ bands exhibit
$e_g$-symmetry and $d_{xy}$ band has
$b_{2g}$-symmetry.\cite{kbmprb} Thus, the peak around 1~eV in the
surface spectra may be attributed to an essentially filled $e_g$
band with $b_{2g}$ band appearing above $E_F$.

In summary, correlation effects are found to be significantly weak
in these 4$d$ systems. This resolves the long standing issue that
arose due to the prediction of large electron correlation in these
systems similar to 3$d$-TMOs. We find that the photoemission and
optical responses, and thermodynamic parameters can be
consistently described for the first time within the first
principle approaches. The first observation of different surface
electronic structure in these weakly correlated 4$d$ systems may
possibly be attributed to the change in symmetry at the surface.

\begin{figure}
\caption{Valence band spectra of (a) SrRuO$_3$ and (b) CaRuO$_3$.
Subtraction of O 2$p$ contributions from He~{\scriptsize II}
spectrum of SrRuO$_3$ is demonstrated in inset (I). The peaks 1
and 2 represent the non-bonding and bonding features. These two
peaks are broadened upto XP resolution to obtain O 2$p$
contributions in the XP spectrum of SrRuO$_3$ as shown in inset
(II). The relative intensity of the features is obtained by least
square error method.}
\end{figure}

\begin{figure}
\caption{(a) TDOS, (b) Ru 4$d$ PDOS and (c) O 2$p$ PDOS in SrRuO$_3$
and CaRuO$_3$. $d_{xy}$, $d_{yz}$ and $d_{xz}$ bands are shown in the
inset in (b) for SrRuO$_3$ and in (c) for CaRuO$_3$.}
\end{figure}

\begin{figure}
\caption{(a) Ru $t_{2g}$ band extracted from He {\scriptsize II}
spectra for CaRuO$_{3}$ and SrRuO$_3$. (b) Ru 4$d$ PDOS convoluted
with Fermi distribution function at 300 K. This represents the
signature of coherent feature. (c) Ru $t_{2g}$ band extracted from XP
spectra for CaRuO$_3$ and SrRuO$_3$. The resolution broadened
He~{\scriptsize II} spectra are also shown for comparison.}
\end{figure}

\begin{figure}
\caption{Surface and bulk spectra in (a) CaRuO$_3$ and (b)
SrRuO$_3$. The solid lines represent the simulated spectra from
first principle calculations. The dashed and dot-dashed lines in
(a) represent the simulated spectra for $U/W$ = 0.12 and 0.36,
respectively. The calculated optical response are shown in the
inset. The lineshape is found to be similar to that observed
experimentally.\cite{ahn} }
\end{figure}


\begin{thebibliography}{99}
\bibitem{note2} A small impurity contribution could not be
avoided as that also observed in the high quality films studied
recently\cite{park}. Reproducibility of the spectra was confirmed
after every scraping cycle.
\end{thebibliography}
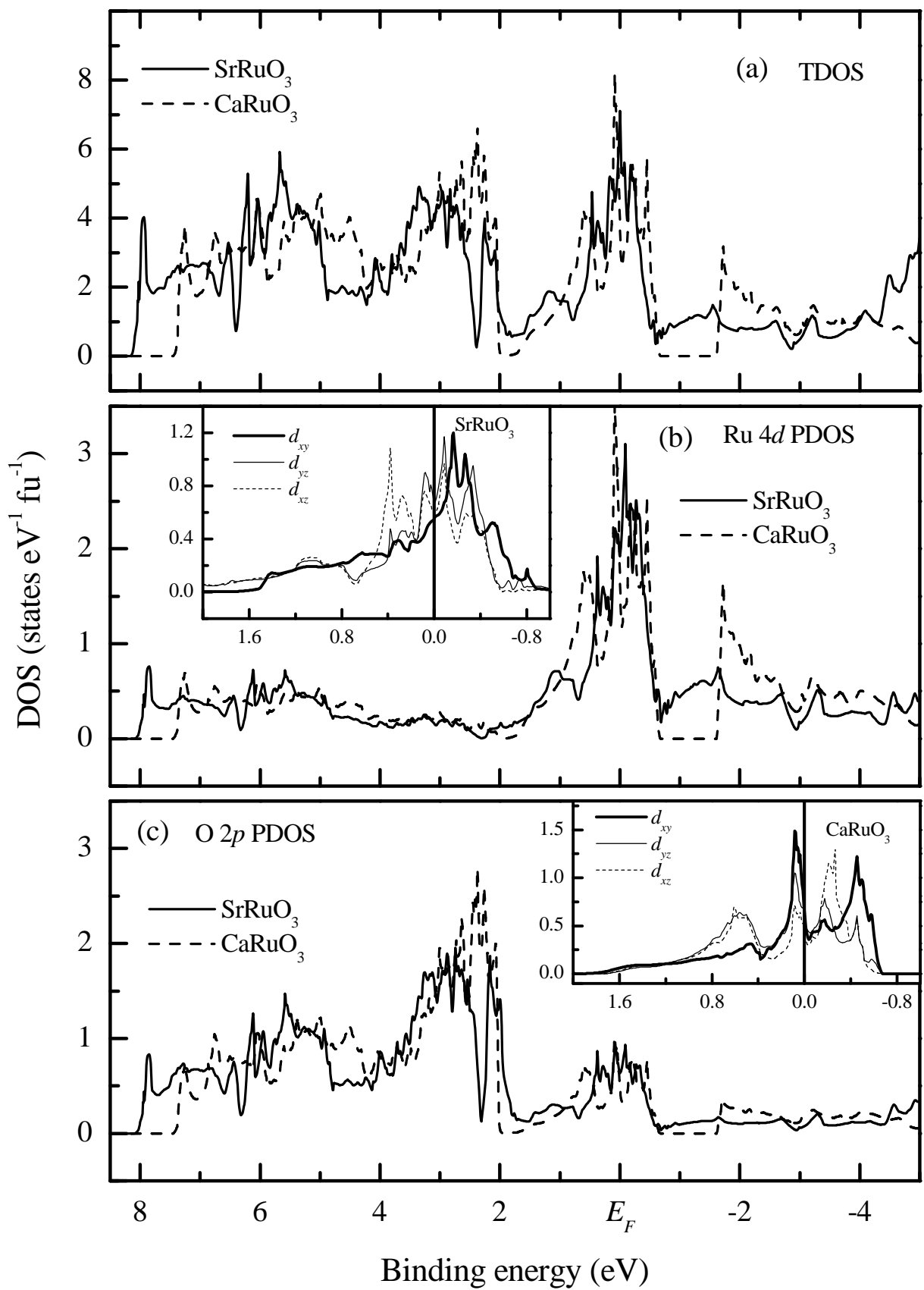
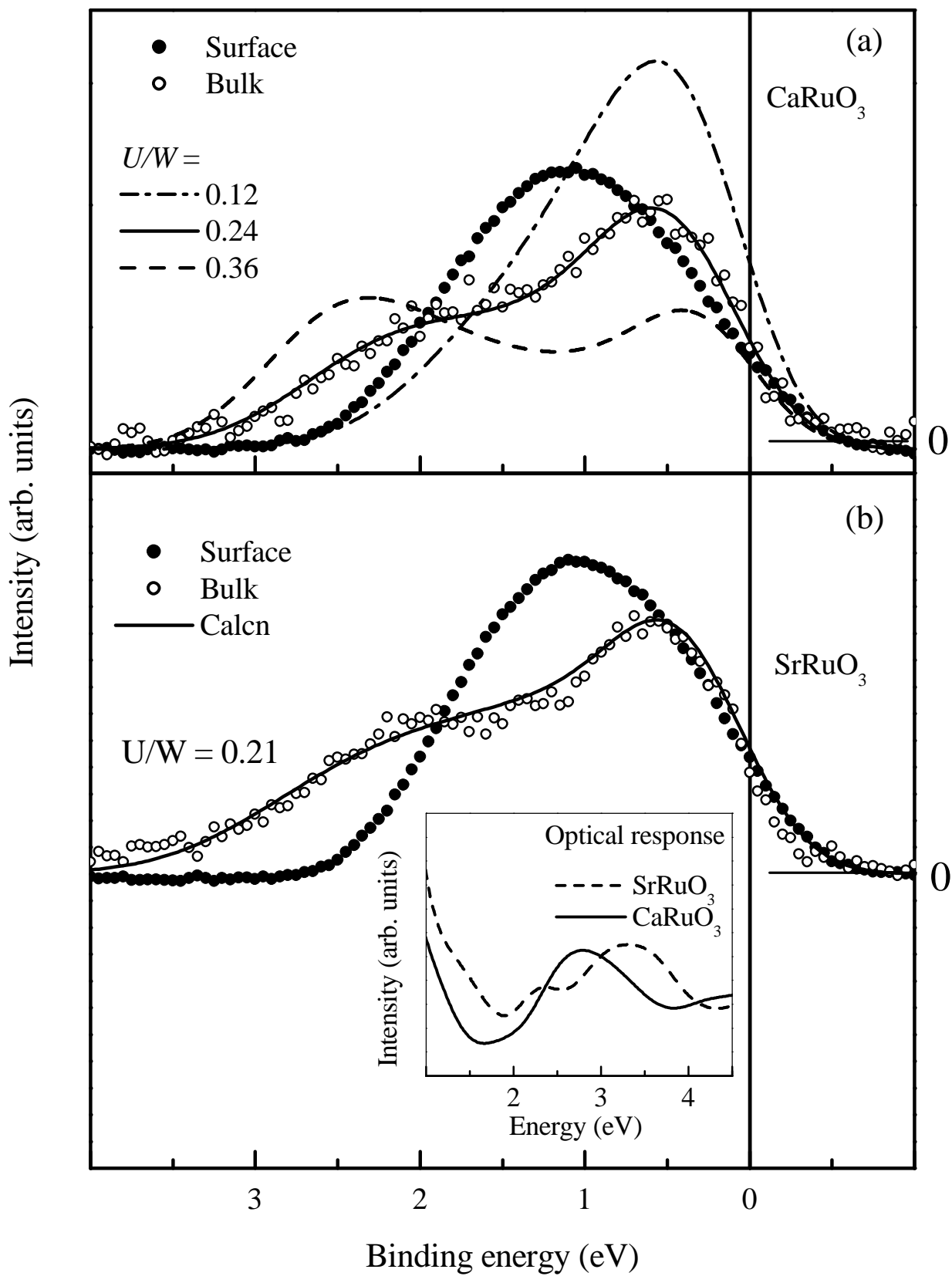
\end{document}